\documentclass[aps,prl,reprint,superscriptaddress]{revtex4-2}

\usepackage{hyperref}
\usepackage{graphicx}
\usepackage{amsmath}
\usepackage{siunitx}
\usepackage{verbatim}
\usepackage{amsfonts}
\usepackage{physics}
\usepackage{todo}

\graphicspath{{figures/}}

\begin{document}

%Title of paper
\title{Enabling the Realisation of Proton Tomography}

\author{B.~T.~Spiers}
\email{benjamin.spiers@physics.ox.ac.uk}
\author{R.~Aboushelbaya}
\author{Q.~Feng}
\author{M.~W.~Mayr}
\author{I.~Ouatu}
\author{R.~W.~Paddock}
\author{R.~Timmis}
\author{R.~H.-W.~Wang}
\affiliation{Department of Physics, Atomic and Laser Physics sub-Department, University of Oxford, Clarendon Laboratory, Parks Road, Oxford OX1 3PU, United Kingdom}
\author{P.~A.~Norreys}
\affiliation{Department of Physics, Atomic and Laser Physics sub-Department, University of Oxford, Clarendon Laboratory, Parks Road, Oxford OX1 3PU, United Kingdom}
\affiliation{Central Laser Facility, UKRI-STFC Rutherford Appleton Laboratory, Harwell Campus, Didcot, Oxfordshire OX11 0QX, United Kingdom} 
\affiliation{John Adams Institute, Denys Wilkinson Building, Oxford OX1 3RH, United Kingdom}

\date{\today}

\sisetup{separate-uncertainty=true,multi-part-units=single}

\begin{abstract}
The proton radiography diagnostic is widely used in laser-plasma experiments to make magnetic field measurements. Recent developments in analysis have enabled quantitative reconstruction of path-integrated magnetic field values, but making conclusions about the three-dimensional structure of the fields remains challenging. In this Letter we propose and demonstrate in kinetic simulations a novel target geometry which makes possible the production of multiple proton beams from a single laser pulse, enabling the application of tomographic methods to proton radiography.
\end{abstract}

\maketitle

%\section{Introduction}

The proton radiography diagnostic allows probing of transient and quasi-static magnetic field structures in plasmas~\cite{borghesi,mackinnon}. It has been used to image magnetic fields in laboratory analogues of astrophysical collisionless shocks~\cite{kugland_2013,morita,park}, the Weibel instability in interpenetrating plasma flows~\cite{huntington}, stochastic magnetic fields amplified by the turbulent dynamo mechanism~\cite{tzeferacos} and fields involved in laser-driven magnetic reconnection~\cite{palmer,tubman}. In the higher-density physics regime, proton radiography has been used to probe plasma fields in imploding inertial fusion capsules in both direct-drive~\cite{rygg,chikang_2009_hed,mackinnon_2006} and indirect-drive laser-hohlraum~\cite{sarri,chikang_2009,chikang_2010} configurations, and in studies of laser channelling physics relevant to fast ignition~\cite{spiers}. Proton-radiographic magnetic field measurements have also been used to validate a Faraday rotation-based magnetic field diagnostic at OMEGA~\cite{rigby}.

\begin{figure}
	\includegraphics[width=\columnwidth]{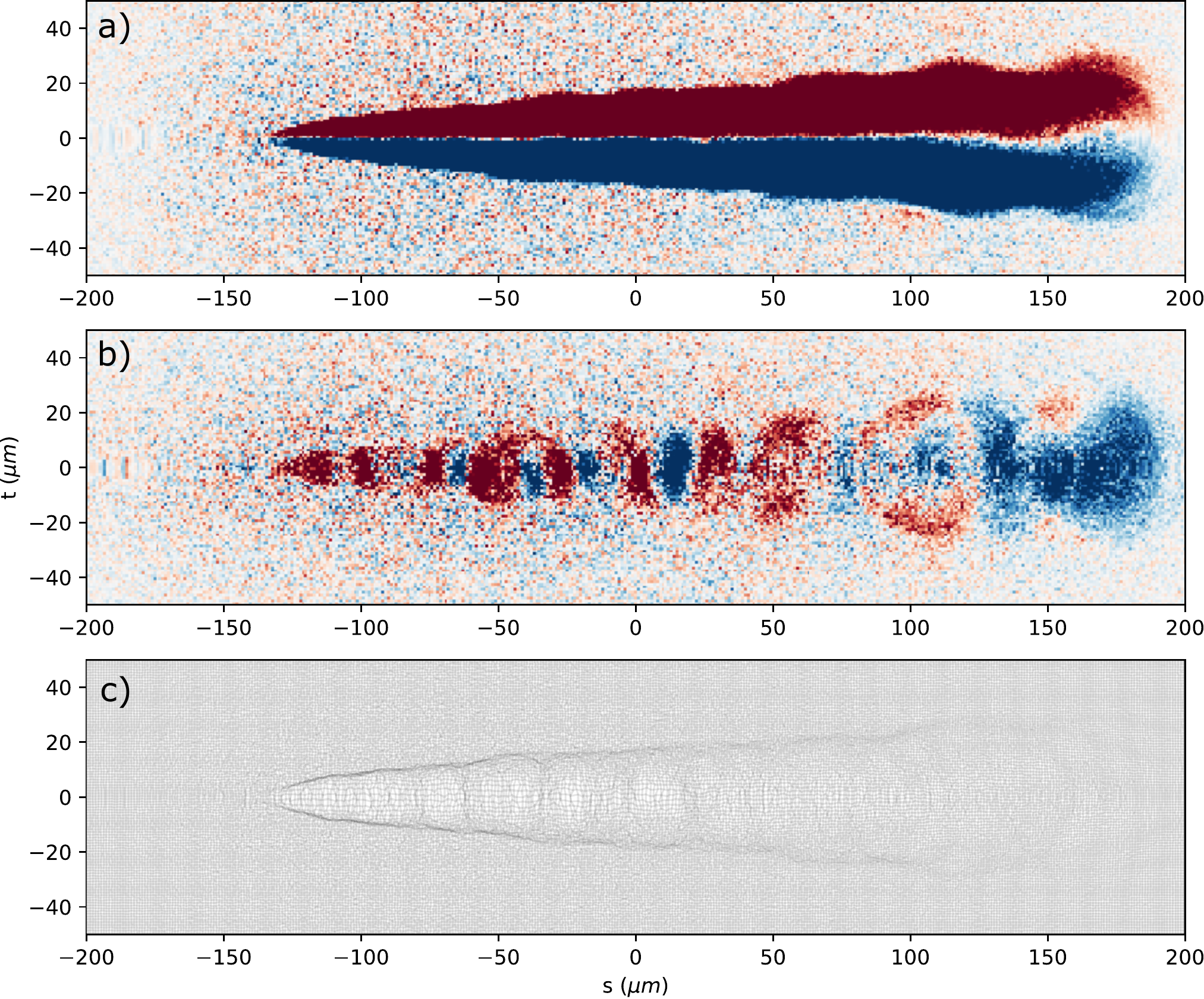}
	\caption{Simulated radiograph of laser channel in plasma. Propagation of a laser pulse into a density gradient was modelled using the particle-in-cell code, \texttt{Smilei}~\cite{smilei}. Path-integrated magnetic field values in the \(s\) and \(t\) directions are shown in panels a) and b) respectively, using a probe direction 20 degrees offset from the laser axis. Deflections of an intially laminar proton beam after traversing are shown in panel c), where darker regions indicate higher proton fluence at the image plane. The \(t\)-component shown in panel b) is the only one accessible to tomography; the \(s\)-component does not transform as a scalar under rotation about \(t\) so cannot be usefully probed by tomographic methods.}
	\label{fig:radiograph}
\end{figure}

An example of the path-integrated fields experienced by a proton beam probing a laser channel at \ang{20} incidence is shown in Fig.~\ref{fig:radiograph}, along with a synthetic radiograph produced using proton deflections corresponding to those fields. \textcite{kugland} presented an early detailed theoretical treatment of proton radiography, followed by proposed inversion techniques by \textcite{kasim1} and \textcite{archie} which detailed algorithms for quantitative recovery of path-integrated magnetic field values, and \textcite{kasim2} addressed uncertainties in the initial source profile used as input to these algorithms. \textcite{chen} investigated machine learning-powered recovery of parametrised field structures from a proton radiograph, and proposed that imaging from several lines of sight could improve the fidelity of such three-dimensional reconstructions. While most proton radiography experiments to date use only one line of sight, \textcite{chikang_2006} have used unusual experimental geometry to probe similar interactions both side-on and face-on in a single shot, and more recently \textcite{tubman} used two proton probes separated by \ang{45} to assist in separating electric and magnetic field contributions to proton deflection. In this Letter's companion article~\cite{companion}, we present the first thorough exploration of this idea, elucidating the capabilities and limitations of the approach we term \emph{proton tomography}. Most significantly we derive two new techniques which can drastically improve the performance of the standard \emph{filtered back-projection} algorithm under conditions expected to be typical for proton tomography.

The first is derived by considering the form of the two-dimensional Fourier transform in polar coordinates, and exploiting filtered back-projection to perform integer-order Hankel transforms. It is shown to be a form of interpolation in the angular variable of the tomographic data domain, which can therefore also be applied to enhance the results of other tomographic inversion algorithms as well as filtered back-projection. This interpolation is shown to radically improve reconstruction quality in the limit of a very small number of probe directions, improving the potential for proton radiography to be implemented practically.

Our second proposed technique is a sampling strategy which improves the efficiency of sampling highly prolate or oblate functions. Many plasma structures which carry distinctive magnetic field signatures commonly probed with proton radiography are highly elongated. Examples include laser-plasma channels, which have been studied using proton radiography by \textcite{spiers} and laboratory analogues to astrophysical relativistic jets such as those studied by \textcite{chikang_2016}. Our derived method allows the user an additional degree of freedom in designing tomographic experiments which can be used to optimise sampling according to the shape of the function.

\begin{figure*}
	\includegraphics[width=\textwidth]{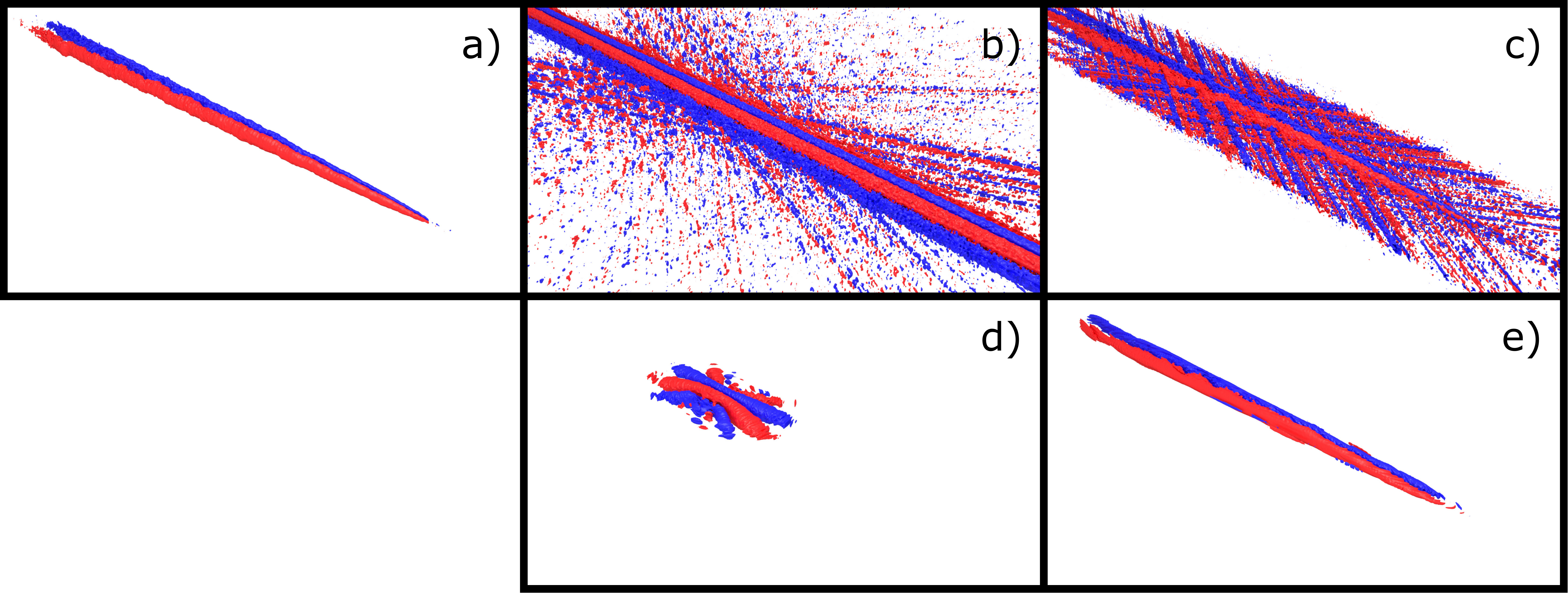}
	\caption{Three-dimensional renderings of a magnetic field structure representing laser-formed channel in plasma, as shown in this Letter's companion article~\cite{companion} (a) and reconstructions from nine projections of the \(B_z\) component (b-e). In c) and e), our aspect ratio compensation method is employed to improve the sampling of the very elongated function. In d) and e), Fourier series interpolation is used to improve the smoothness of the reconstruction by removing artefacts arising from the sparse angular sampling rate. It is clear that in this regime - sparse sampling of an elongated function - only the combination of both enhancements results in a good reconstruction of the original function. Further testing has shown that the method is not highly sensitive to the precise aspect ratio used in the compensation, with ratios of 20:1 and 5:1 showing similar performance to the 10:1 compensation shown here. All functions are rendered as isosurfaces calculated at \(\pm\SI{1}{\kilo\tesla}\), with the positive isosurface shown in red and the negative in blue.}
	\label{fig:reconstructions}
\end{figure*}

An example of a tomographic reconstruction of a three-dimensional magnetic field structure associated with a laser channelling into plasma is shown in Figure~\ref{fig:reconstructions} demonstrating the failure of standard methods as well as the improvement that can be achieved using our proposed modifications.

While the number of proton beams which can be simultaneously produced on a single shot of a high-powered laser system will be a limiting factor for reconstruction quality, in this Letter we propose a novel experimental target geometry which enables production of several proton beams with independently controllable pointing from a single laser-foil interaction. A simplified version of this target geometry is validated in two-dimensional particle-in-cell simulations, producing two independent proton beams from a single laser interaction. This method is easily extended to drive more wire-foil components from a single short pulse. For example, a version driving four proton beams per short pulse could be implemented at a system such as the National Ignition Facility's Advanced Radiographic Capability (NIF-ARC), whose four short pulse beamlets would make available up to sixteen simultaneous proton beams---more than the number used to produce the results of Figure~\ref{fig:reconstructions}---which would greatly improve the ability of experimenters to diagnose complicated three-dimensional field structures within the target.

%\section{Methods}
%\label{sec:methods}

\begin{figure}
	\includegraphics[width=0.8\columnwidth]{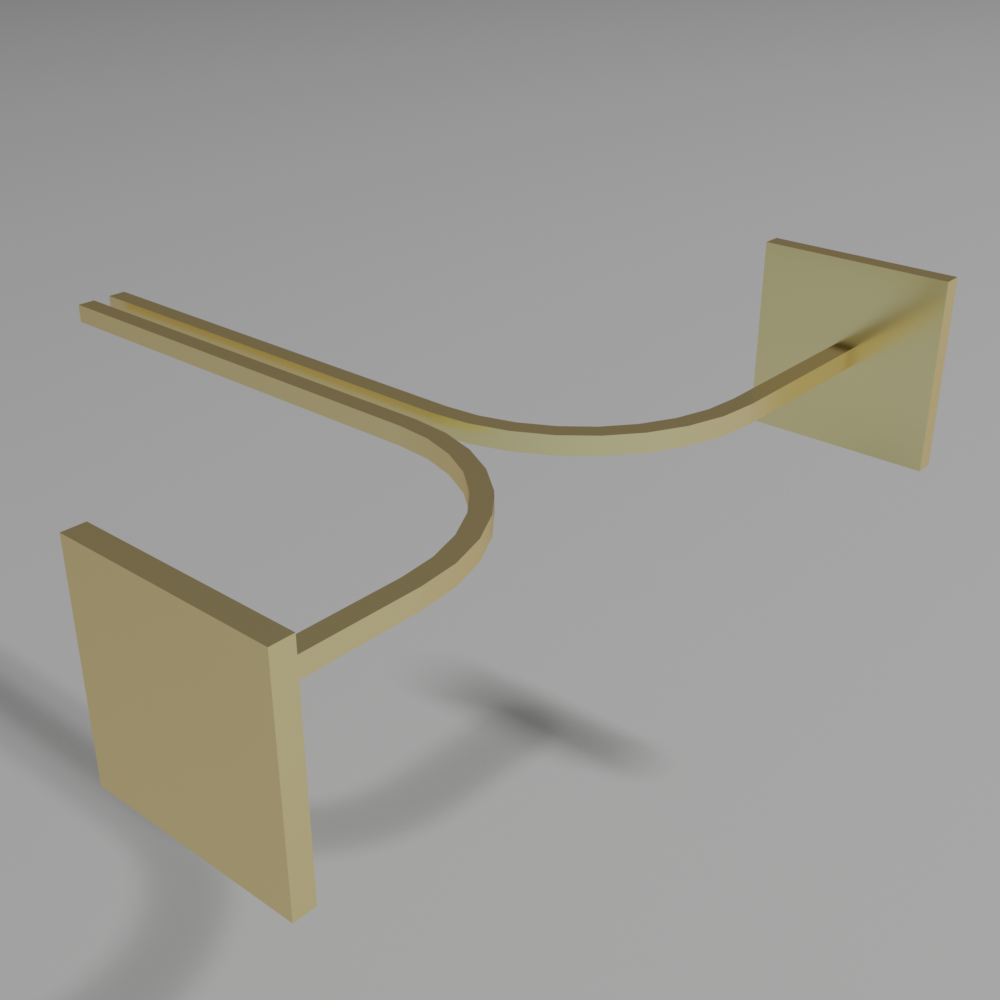}
	\caption{Three-dimensional rendering of the proposed target geometry, as tested in particle-in-cell simulations. A \SI{10}{\micro\metre}-thick gold foil (the `laser' foil) is attached by \SI{10}{\micro\metre} diameter wires to two smaller `beam' foils of the same thickness. The `laser' foil is struck from the rear by a short, high-contrast laser pulse and protons present as impurities in the `beam' foils are diagnosed when they leave the simulation box.}
	\label{fig:target}
\end{figure}

The proposed scheme consists of several metal foils connected to conductive wires. The foils are located at desired points in space, according to the requirements of experimental geometry, and the conductive wires connected to their rear faces are bundled together. When a laser pulse is incident on the bundled ends of the wires, electrons are expelled and a space-charge imbalance is set up in and around the wire ends. This outflow of electrons is balanced by large return currents along the wires' surfaces. These currents ultimately lead to positive charging of the secondary foils, expelling the relatively light hydrogen ions while the gold ions respond more slowly.

Two-dimensional particle-in-cell simulations were run using the code \texttt{Smilei}~\cite{smilei}. The target was initialised with density profile given by the `test' target geometry rendered in Fig.~\ref{fig:target}, and was composed of gold ions, electrons and a smaller density of hydrogen ion impurities. Protons were initialised using two separate species definitions: the first in the wires, and the second in the two beam foils. This allowed protons from the beam foils to be distinguished in diagnostics and also allowed them to be modelled using a higher particle-per-cell count than in the rest of the simulation, in order to improve the statistical properties of the measured ion spectra. Additional simulations were run using a `classical' target-normal sheath accleration target configuration, with protons accelerated from a single, larger foil measured to enable comparison of the spectra produced by `classical' target-normal sheath acceleration to the spectra of ions accelerated from the foils in the full simulation set-up.

\begin{figure}
	\includegraphics[width=0.8\columnwidth]{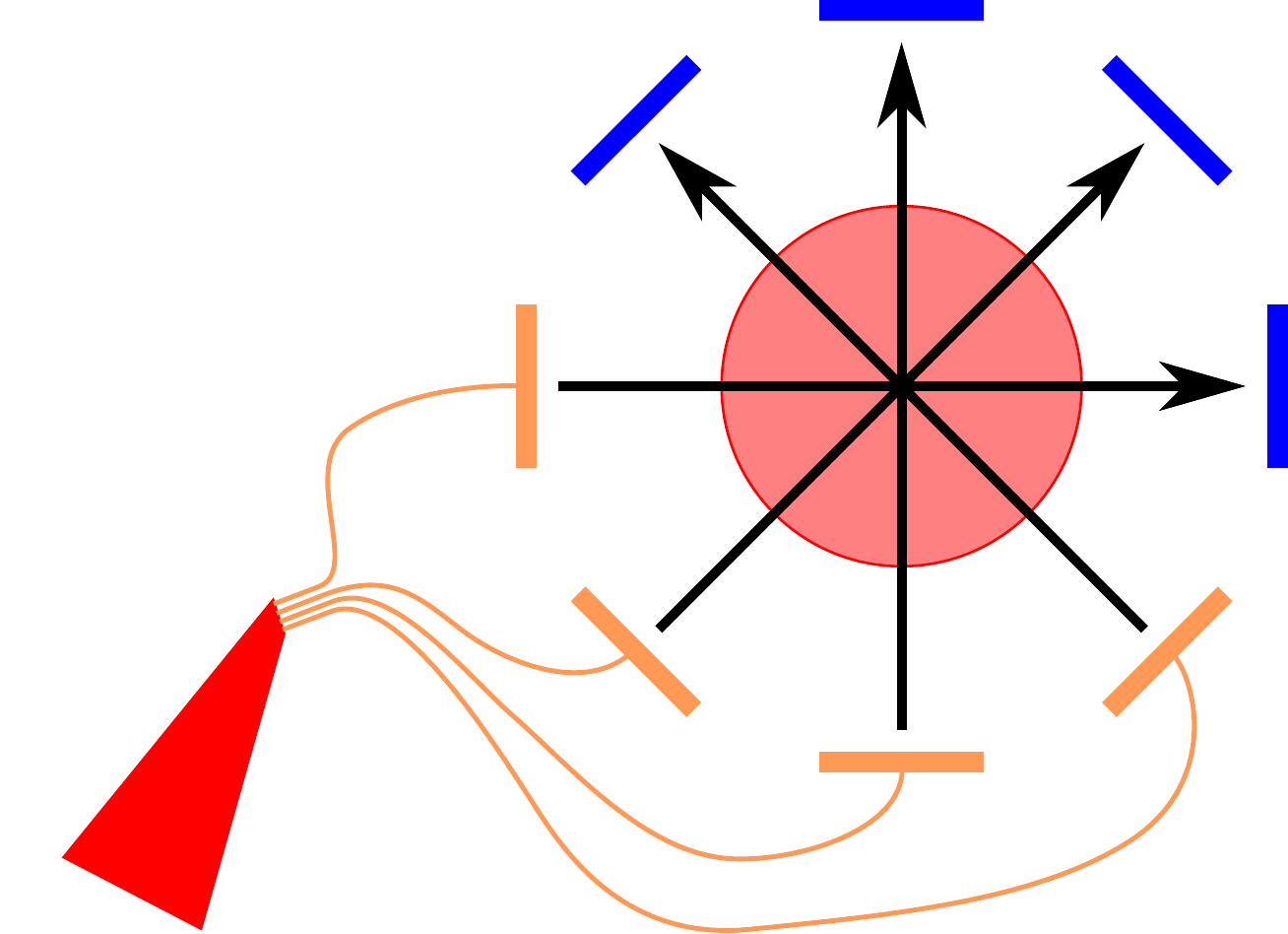}
	\caption{The proposed scheme used to probe a single target with four proton beams. As discussed above the use of multiple short pulses would enable more proton beams to be driven simultaneously, but for simplicity this schematic only shows one short pulse driving a bundle of four beam foils.}
	\label{fig:tomo}
\end{figure}

Figure~\ref{fig:tomo} illustrates the proposed use of this new target geometry in schematic form. This includes four foil-wire arrangements driven by a single laser pulse rather than two, with wires covering larger distances than would be practical to simulate in a particle-in-cell framework.

%\section{Results}
%\label{sec:results}

\begin{figure*}
	\includegraphics[width=\textwidth]{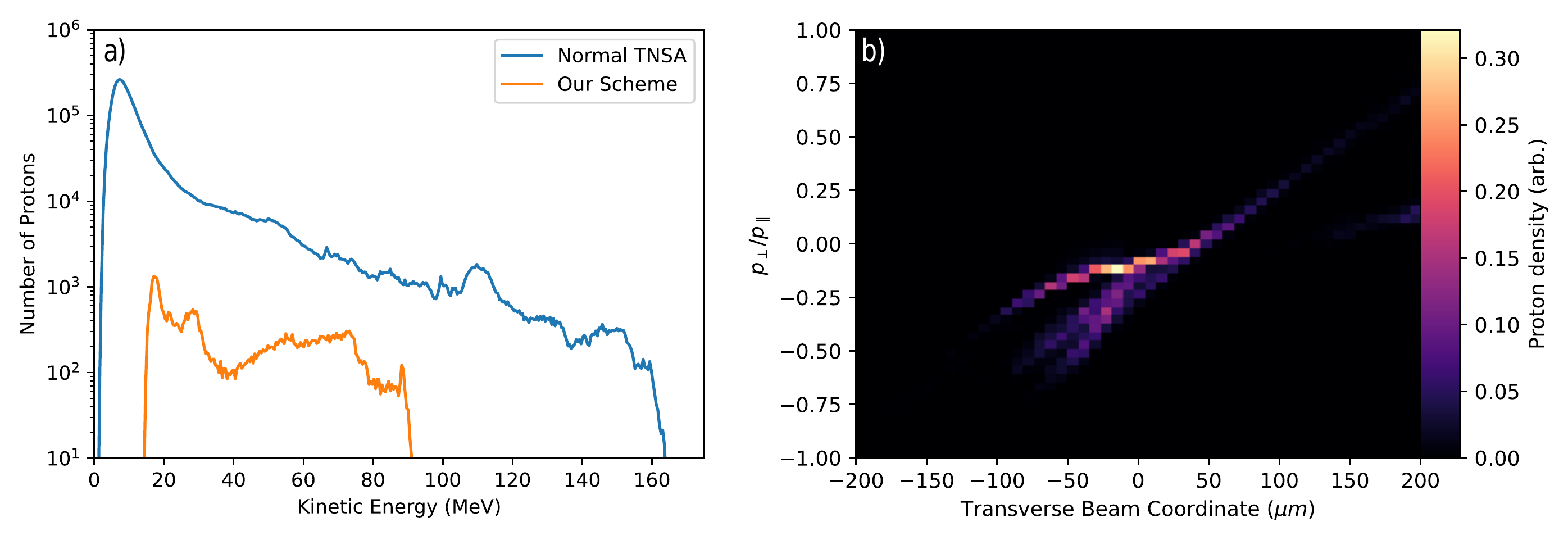}
	\caption{a) Spectra of ions accelerated in a `classical' TNSA configuration and for ions accelerated from the foils in the proposed scheme. Proton energies are recorded as they leave the simulation through the positive y boundary. b) Laminarity diagnostic of protons acceelerated in the new scheme. A perfectly laminar beam produced from a point source a distance \(z_s\) behind the measurement plane would fulfil the condition \(p_\perp / p_\parallel = x_\perp / z_s\); the resemblance of panel b) to a straight line indicates a good degree of laminarity.}
	\label{fig:protonspectrum}
\end{figure*}

In Figure~\ref{fig:protonspectrum} the proton number spectrum as a function of kinetic energy is shown for both a single foil, i.e. the `classical' TNSA arrangement~\cite{mackinnon_2001}, and for ions accelerated from a secondary foil in our proposed set-up. The spectrum produced by TNSA has the expected, exponential energy spectrum, with a cut-off energy around \SI{160}{\mega\electronvolt} and temperature of approximately \SI{35}{\mega\electronvolt}. Contrastingly, the spectrum produced from the foils in the proposed scheme has significantly narrower energy distribution, in the range \SIrange{15}{90}{\mega\electronvolt}, and appears somewhat more uniform than the TNSA spectrum, albeit at the cost of an order of magnitude fewer accelerated protons per unit energy. A histogram over the transverse coordinate and propagation direction of protons accelerated by the new scheme is also shown, demonstrating the desirable laminarity properties of the resulting beam.

%\section{Conclusion}
%\label{sec:conclusion}

The novel target construction presented in this Letter has been shown to allow more flexible design of proton radiography diagnostics, by enabling a single laser pulse to produce several independently pointed proton beams. This has been validated using two-dimensional particle-in-cell simulations, which showed that the spectrum of protons accelerated from these `secondary' foils agrees well with that of ions accelerated from the `primary' foil in the absence of additional components. We have argued that this target design is feasible for fielding on large laser systems featuring short-pulse capabilities, such as NIF-ARC, OMEGA EP and LMJ-PETAL. In the first of these cases, the Advanced Radiographic Capability at the NIF consists of four beamlets which may be pointed independently, and we envision using each of these four beamlets to drive one primary and four secondary foils~\cite{nifarc,mariscal}. This provides sixteen proton beams with similar properties. This is in greater than the number of projections shown in Figure~\ref{fig:reconstructions} to produce a good reconstruction of a laser-driven magnetic field structure when both the Fourier interpolation and aspect ratio compensation schemes derived in the companion article~\cite{companion} are used, though both are necessary for good reconstruction quality. This new proposed experimental capability therefore represents a major step towards the realisation of fully three-dimensional magnetic field reconstruction in laser-plasma experiments.

\begin{acknowledgments}
The authors gratefully acknowledge the support of all the staff at the Central Laser Facility, UKRI-STFC Rutherford Appleton Laboratory and the ORION Laser Facilty at AWE Aldermaston (particularly discussions with Gavin Crow) while undertaking this research. This work has been carried out within the framework of the EUROfusion Consortium and has received funding from the Euratom research and training programme 2019-2020 under grant agreement No 633053, with support of STFC grant ST/P002048/1 and EPSRC grants EP/R029148/1 and EP/L000237/1. The views and opinions expressed herein do not necessarily reflect those of the European Commission. BTS acknowledges support from UKRI-EPSRC and AWE plc. PAN acknowledges support from OxCHEDS for his William Penney Fellowship. The simulations presented herein were carried out using the ARCHER2 UK National Supercomputing Service.
\end{acknowledgments}

\bibliography{prldraft}

\end{document}